\documentclass[12pt,a4paper]{article}

  \usepackage{a4wide}
  \usepackage{latexsym}
  \usepackage{epsf}
  \usepackage{amssymb}
  \usepackage{graphicx}
  \usepackage{amsmath, cite}
  \usepackage{slashed,epsfig}
  \usepackage{bbm}
  \usepackage{bbm}
  \usepackage{amsmath,amssymb,amsthm}
  \usepackage{pst-all}
  \usepackage{here}
  \usepackage{hyperref}

\renewcommand{\d}{\textrm{d}}
\newcommand{\Real}{\textrm{I\!R}}
\def\rme{{\mathrm e}}
\newcommand{\e}{\textrm{e}}
\renewcommand{\d}{\textrm{d}}

\newcommand{\SO}{\mathop{\rm SO}}

\newcommand{\SL}{\mathop{\rm SL}}

\begin{document}

\begin{flushright}
\small
UUITP-29/10\\
UG-2010-76\\
\date \\
\normalsize
\end{flushright}

\begin{center}

\vspace{.7cm}

{\LARGE {\bf  Black holes  as  generalised Toda molecules }} \\

\vspace{1.2cm}

{\large Wissam Chemissany$^a$, Jan Rosseel$^{b}$ and  Thomas Van Riet$^c$} {}~\\
\vspace{1cm}

$^a$ {\small\slshape University of Lethbridge, Physics Dept.,
Lethbridge T1K 3M4, Alberta, Canada }\\
{\upshape\ttfamily wissam.chemissany@uleth.ca}\\\vspace{0.2cm}

$^b$ {\small\slshape Centre for Theoretical Physics, University of Groningen,\\
    Nijenborgh 4, 9747 AG Groningen, The Netherlands}\\
{\upshape\ttfamily J.Rosseel@rug.nl}\\
    \vspace{0.2cm}

$^c$ {\small\slshape Institutionen f\"{o}r Fysik och Astronomi, Box 803, SE-751 08 Uppsala, Sweden} \\
{\upshape\ttfamily thomas.vanriet@fysast.uu.se}

\vspace{3cm}

{\bf Abstract} \end{center} {\small In this note we compare the
geodesic formalism for spherically symmetric black hole solutions
with the black hole effective potential approach.  The geodesic
formalism is beneficial for symmetric supergravity theories since
the symmetries of the larger target space leads to a
complete set of commuting constants of motion that establish the
integrability of the geodesic equations of motion, as shown in
\href{http://arxiv.org/abs/1007.3209}{arXiv:1007.3209}.  We point
out that the integrability lifts straightforwardly to the
integrability of the equations of motion with a black hole potential.
 This construction turns out to be a
generalisation of the connection between Toda molecule equations and
geodesic motion on symmetric spaces known in the mathematics
literature. We describe in some detail how this generalisation of
the Toda molecule equations arises.}

\newpage

\pagestyle{plain} \tableofcontents
\section{Introduction}
To find solutions in supergravity theories is known to be
complicated due to the non-linear structure of the Einstein
equations and the complexity of having many fields, especially
scalar fields. However, when solutions have enough space-time
symmetries there is often only a single coordinate that the fields
depend on. For cosmological solutions this is the time coordinate
and for spherical stationary black hole solutions this is the radial
distance from the singularity. This dependence on a single
coordinate implies that the equations of motion reduce to ordinary,
but coupled, second-order differential equations of the kind we
encounter in Hamiltonian systems. In the specific case of black hole
solutions\footnote{From now on we assume the black hole solutions to
be stationary and spherically symmetric.} in \emph{massless}
supergravity theories there are two known ways to describe these
Hamiltonian systems. The first way, originally pioneered in
\cite{Breitenlohner:1987dg} uses the timelike Killing vector to
reduce the problem to a supergravity problem in three Euclidean
dimensions. In three dimensions all vectors can be dualised to
scalars, such that we have gravity coupled to some non-linear sigma
model. It can be shown that the Einstein equations become trivial
and that the scalar field equations of motion reduce to the
equations of motion for a geodesic curve on the sigma model. The
inverse of the radial coordinate $r$ turns out to give an affine
parametrisation of the geodesic. Hence the effective action can be
written as ($\tau=1/r$)
\begin{equation}
S_a=\int\d\tau\Bigl(
\tfrac{1}{2}\,g_{ij}\dot{\phi}^i\dot{\phi}^j\Bigr)\,,
\end{equation}
where $g_{ij}$ is the sigma model metric in three-dimensional
supergravity. This metric is indefinite due to the reduction over a
timelike direction \cite{Breitenlohner:1987dg, Hull:1998br,
Bergshoeff:2008be, Cortes:2009cs}.

The second method for constructing an effective action, which we
name $S_b$, was developed in \cite{Ferrara:1995ih, Ferrara:1997tw}
and uncovered the well-known attractor mechanism for some extremal
black holes. The construction of $S_b$ works when we assume that the
NUT charge vanishes and the stationarity turns into staticity. Then
one can integrate out the vector fields in four dimensions in terms
of their electric and magnetic charges. The remaining equations of
motion can then be derived from the following one-dimensional action
\begin{equation}
S_b=\int\d\tau \Bigl(\dot{U}^2
+\tfrac{1}{2}G_{rs}\dot{\phi}^r\dot{\phi}^s -
\e^{2U}V_{BH}(\phi)\Bigr)\,,
\end{equation}
where $U$ is the so-named black hole warp factor which describes the
four-dimensional space-time metric (see below). The scalars $\phi^r$
are the scalars of the four-dimensional supergravity theory and span
a submanifold (together with $U$) of the sigma model in three
dimensions. The function $V_{BH}(\phi)$ is named the black hole
effective potential and contains the electric and magnetic charges.
We re-derive both effective actions in section 2 of this paper. One
can think of the term $\e^{2U}V_{BH}$ as a scalar potential in two
dimensions originating from a flux compactification over the
two-sphere.

In this paper we specify to the case of ``symmetric supergravity
theories''. By that we mean that the scalars span a symmetric coset
space $G_D/H_D$, where by the subscript $D$ we denote the dimension of the spacetime of the theory.  Symmetric supergravity theories that are obtained from dimensionally reducing over time have an indefinite signature of the scalar manifold. This implies that $H$ is not the maximal compact subgroup of $G$, but some non-compact subgroup.

It is in these circumstances that the geodesic approach seems
beneficial since we have a larger symmetry group $G_3 \supset G_4$.
This larger symmetry comes with the following concrete benefits:
\begin{itemize}
\item The geodesic equations of motion can be recast in a simple
form in terms of the symmetric coset matrix $\mathcal{M}$. The
solution to these equations is a simple matrix exponential
$\mathcal{M}(\tau)=\mathcal{M}(0)\e^{Q\tau}$
\cite{Breitenlohner:1987dg}.
\item Solutions can be generated using the large symmetry group
$G_3$ on simple ``seed solution'', see \cite{Breitenlohner:1987dg,
Gal'tsov:1998yu,Bergshoeff:2008be} (and \cite{Chemissany:2007fg} for
a simple pedagogical example). The $G_3$-orbit structure gives the
essential insight for the understanding of regularity and
supersymmetry of the various solutions \cite{Breitenlohner:1987dg,
Gaiotto:2007ag, Bergshoeff:2008be, Bossard:2009at}\footnote{See also
\cite{Cortes:2009cs} for a general treatment of
dimensionally-reduced black holes in supergravity theories that are
not necessarily symmetric.}. \item The geodesic equations of motion
also allow a Lax pair form \cite{Chemissany:2009hq, Fre:2009et,
Chemissany:2009af}, of a kind that allows a closed but iterative
formula for the coset representative $L$. Because of the
upper-triangular structure of the coset representative $L$ in the
Borel gauge this allows to peal of the solutions for the individual
scalar fields $\phi^i$, something that is more complicated for the
matrix $\mathcal{M}$. A different, but equivalent Lax pair formalism
can be found in \cite{Figueras:2009mc}.

\item The Lax pair formalism has been used to prove the full
Liouville integrability of the geodesic equations of motion. This
means that for an $n$-dimensional symmetric space one can find $n$
constants of motion that mutually Poisson commute
\cite{Chemissany:2010zp}. This in turn implies that the system can
\emph{always} be integrated in the sense of Hamilton--Jacobi
\cite{Chemissany:2010zp}. This has been called the `` first-order
formalism '' or the `` fake supergravity '' formalism in the
supergravity literature, see \cite{Ceresole:2007wx,
LopesCardoso:2007ky, Andrianopoli:2007gt, Janssen:2007rc,
Andrianopoli:2009je, Bossard:2009we, Andrianopoli:2010bj,
Ceresole:2010nm} and references therein.
\item Finally, we refer to \cite{Mohaupt:2009iq,Mohaupt:2010fk} for
recent applications of the geodesic approach in understanding
properties of black hole solutions, such as the defining harmonic
functions, the attractor structure and a canonical way of obtaining
non-extremal solutions from extremal ones. We refer to
\cite{Levay:2010ua, Borsten:2010db} for an interesting connection
with quantum information theory\footnote{ Much more work on
uncovering the structure of supergravity solutions has been carried
out in the recent literature using the geodesic approach and we
refer the interested reader to the references in the cited papers
for further reading.}.
\end{itemize}

It is the aim of this paper to clarify how some of the benefits of
the geodesic approach can also be made visible in the black hole
effective potential action, especially the issue regarding the
integrability. Outside of the black hole context this was
investigated by mathematicians in the context of Toda molecule
equations. In \cite{Olshanetsky:1981dk} (see also
\cite{Ferreira:1984bi}) it was established that the Toda molecule
equations, which are a Hamiltonian system with specific potential,
could be viewed as having an underlying geodesic motion on an
enlarged target space. Even more, it is conjectured in
\cite{Olshanetsky:1981dk} \emph{that all one-dimensional integrable
systems have an underlying geodesic motion on a symmetric space}. We
consider it to be very interesting that this is realised in the
context of black holes in supergravity theories as the equivalence
between the two approaches (geodesic and black hole potential) to
construct the effective action. As we demonstrate in detail below,
the connection between the two approaches generalises the link
between Toda molecules and symmetric spaces in two ways: i) we find
generalised Toda molecule equations from truncating the geodesic
motion in a more general way then done in \cite{Olshanetsky:1981dk,
Ferreira:1984bi} and ii) we point out that there exist different
choices for the signature of the sigma model without ruining the
integrability, which also defines a generalisation of Toda molecule
equations.

This paper is organised as follows. In section
\ref{effectiveactions} we rederive the two known effective actions
for general theories of gravity coupled to massless scalars and
Abelian gauge fields. In section \ref{symmetric} we specify to
symmetric supergravity theories and first review the construction of
the geodesic motion and its integrability. Then we point out in
section \ref{integrability} how this integrability implies, in a
straightforward manner, the Liouville integrability of the action
with potential. In section \ref{todamolecules} we describe the link
with Toda molecule equations and discuss the generalisation that
black hole effective actions naturally provide. Finally in section
\ref{discussion} we end with a discussion and a conjecture about
integrability of domain wall and cosmological solutions in gauged
supergravity.

\section{Effective actions}\label{effectiveactions}
In this section we review the effective actions for spherically
symmetric and stationary black hole solutions in theories with
multiple Abelian gauge fields  $B^I$, $I=1,\dots, n_V$ and scalar
fields $\phi^r$ (without mass terms). This has been reviewed in many
papers\footnote{We refer to \cite{VanProeyen:2007pe} for a
construction of black hole effective actions in a more general
setting.} but here we wish to present this again in order to
establish our conventions and notation but also because the
comparison between the effective actions is our main focus.

The action in four dimensions is given by\footnote{Our discussion
trivially extends to any dimension. The existence of electric and
magnetic point charges is however specific to four dimensions.}
\begin{equation}\label{4Daction}
S_4=\int\Bigl(\tfrac{1}{2}\star R_4 - \tfrac{1}{2}G_{rs}\star\d
\phi^r\wedge\d\phi^s-\tfrac{1}{2}\mu_{IJ}\star G^I\wedge
G^J+\tfrac{1}{2}\nu_{IJ}G^I\wedge G^J\Bigr)\,,
\end{equation}
where $G^I=\d B^I$, and $G_{rs}$, $\mu_{IJ}$, $\nu_{IJ}$ are
symmetric matrices that depend on the scalars $\phi$; in particular,
$G$ and $\mu$ are required to be positive definite. We closely
follow the appendix of \cite{Chemissany:2010zp}, which itself is
based on the original paper by Breitenlohner, Gibbons and Maison
\cite{Breitenlohner:1987dg}.

The Ansatz for black hole solutions can be written as
\begin{align}
& \d s_4^2 =-\e^{2U}(\d t + A_{KK})^2 + e^{-2U}\d s_3^2\,,\nonumber\\
& B^I= \tilde{B}^I + Z^I(\d t + A_{KK})\,,\label{reductionAnsatz}
\end{align}
where $\tilde{B}^I$ and $A_{KK}$ are vectors  and $U$, $Z^I$ are
scalar fields in $d=3+0$. When $A_{KK}$ can not be redefined away
($\d A_{KK}\neq 0$) the black hole has a non-zero NUT charge and is
not static anymore but stationary. We have written the Ansatz in the
same way as a Kaluza--Klein reduction over time. For a true
Kaluza--Klein reduction we would require time to be periodic and
this would manifest itself in the Kaluza--Klein tower being discrete
instead of continuous. Since we restrict to the zero modes this
difference is of no importance.

The Ansatz for the spatial (3-dimensional) part of the metric is
\begin{equation}\label{metric1}
\mathrm{d}s^2_3=\exp[4A(\tau)]\mathrm{d}\tau^2 +
\exp[2A(\tau)]\mathrm{d}\Omega_2^2\,,
\end{equation}
where $\d\Omega_2^2$ is the metric on the unit 2-sphere. Consistency
with the metric symmetries requires the scalar fields to depend on
$\tau$ only, $\phi^r=\phi^r(\tau)$.

Let us now reduce the action over the timelike direction
\begin{align} S_3&=\int\Bigl(\tfrac{1}{2}\star R_3 - \star\d
U\wedge\d U +\tfrac{1}{4}\e^{4U}\star F_{KK}\wedge F_{KK} -
\tfrac{1}{2}G_{rs}\star\d \phi^r\wedge\d\phi^s \nonumber \\ & +
\tfrac{1}{2}\mu_{IJ}\e^{-2U}\star \d Z^I\wedge \d Z^J
-\tfrac{1}{2}\e^{2U}\mu_{IJ}\star(\tilde{G}^I+Z^IF_{KK})\wedge
(\tilde{G}^J+Z^JF_{KK})\nonumber
\\ & -\nu_{IJ}(\tilde{G}^I+Z^IF_{KK})\wedge \d Z^J\Bigr)\,,
\end{align}
where $\tilde{G}^I=\d\tilde{B}^I$, $F_{KK}=\d A_{KK}$. As usual the
vectors $A_{KK}$ and $\tilde{B}^I$ can be dualised to scalars $\chi$
and $Z_I$ by adding the following Lagrange multipliers to the action
\begin{equation}
S'_3=S_3 +\chi \d F_{KK} + Z_I\d \tilde{G}^I\,.
\end{equation}
Varying the action $S_3'$ with respect to $F_{KK}$ and $\tilde{G}^I$
gives the equations of motion
\begin{align}
\d Z_J  &=-\e^{2U}\star \mu_{IJ}(\tilde{G}^I+Z^IF_{KK})-\nu_{IJ}\d Z^I \,,\label{dual1}\\
\d \chi &= \tfrac{1}{2}\e^{4U}\star F_{KK} +Z^I\d
Z_I\,.\label{dual2}
\end{align}
Dualisation of the action $S_3$ is obtained by eliminating $F_{KK}$
and $\tilde{G}^I$ from  the action $S_3'$ using (\ref{dual1},
\ref{dual2}). If we furthermore define, $2\chi \equiv a+Z^IZ_I$ we
find
\begin{align}
S_3&=\int\Bigl(\tfrac{1}{2}\star R_3 - \star\d U\wedge\d U
-\tfrac{1}{2}G_{rs}\star\d \phi^r\wedge\d\phi^s +\tfrac{1}{2}\e^{-2U}\star \d {\bf Z}^T \wedge \mathcal{M}_4 \d {\bf Z} \nonumber\\
& -\tfrac{1}{4}\e^{-4U}\star(\d a + {\bf Z}^T \mathbb{C}\d {\bf
Z})\wedge (\d a + {\bf Z}^T \mathbb{C}\d {\bf
Z})\Bigr)\,,\label{sigma3D}
\end{align}
where
\begin{equation}
{\bf Z}\equiv (Z^I, Z_I)\,,\qquad
\mathbb{C}=\begin{pmatrix}0 & -\mathbbm{1}\\
+\mathbbm{1} & 0 \end{pmatrix}\,,\qquad \mathcal{M}_4 =
\begin{pmatrix} \mu
+\nu\mu^{-1}\nu & \nu\mu^{-1}\\
\mu^{-1}\nu & \mu^{-1}
\end{pmatrix}\,.
\end{equation}
As explained in many papers (see e.g. \cite{Janssen:2007rc}) the
3-dimensional Einstein equations for $A(\tau)$ decouple and are
trivial to solve. The main task is to solve the equations for the
scalars $U, \phi^r$ and ${\bf Z}$. The latter equations can be
derived from the one-dimensional geodesic action, denoted $S_a$,
\begin{equation}\label{S_a}
\boxed{ S_a=\int\d \tau \,\,\Bigl(\dot{U}^2+
\tfrac{1}{2}G_{rs}\dot{\phi}^r\dot{\phi}^s -
\tfrac{1}{2}\e^{-2U}\dot{{\bf Z}}^T \mathcal{M}_4 \dot{{\bf Z}} +
\tfrac{1}{4}\e^{-4U}(\dot{a} + {\bf Z}^T \mathbb{C}\dot{{\bf
Z}})^2\Bigr)\,,}
\end{equation}
where a dot denotes differentiation with respect to $\tau$.

Let us now consider the NUT charge to be zero. It can be shown that
this implies the consistent truncation
\begin{equation}
\dot{a} + {\bf Z}^T \mathbb{C}\dot{{\bf Z}}=0\,.
\end{equation}
Then one observes that the axions $\bf{Z}$ appear shift-symmetric in
the action such that they can be integrated out in terms of the
physical electric and magnetic charges
\begin{equation}
{\bf Q}= (m^I, e_I)\,,
\end{equation}
as follows
\begin{equation}\label{integratingout}
{\bf Q}=\e^{-2U}\mathbb{C}\mathcal{M}_{4} \dot{{\bf Z}}\,.
\end{equation}
When we plug this back into the action we have to add an overall
minus sign in front of the ${\bf Z}$-kinetic term to obtain an
effective action that leads to the correct equations of motion. If
we furthermore use that $\mathcal{M}_4$ is a symplectic matrix
\begin{equation}
\mathcal{M}_{4}^{-1}=\mathbb{C}\mathcal{M}_{4}\mathbb{C}^T\,,
\end{equation}
we find
\begin{equation}
S_3=\int\Bigl(\tfrac{1}{2}\star R_3 - \star\d U\wedge\d U
-\tfrac{1}{2}G_{rs}\star\d \phi^r\wedge\d\phi^s
-\tfrac{1}{2}\e^{2U}{\bf Q}^{T}\mathcal{M}_{4}{\bf Q} \Bigr)\,.
\end{equation}
The one-dimensional effective action $S_b$ reads
\begin{equation}\label{S_b}
\boxed{ S_b=\int\d \tau \,\,\Bigl(\dot{U}^2+
\tfrac{1}{2}G_{rs}\dot{\phi}^r\dot{\phi}^s + \tfrac{1}{2}\e^{2U}{\bf
Q}^{T}\mathcal{M}_4{\bf Q}\Bigr)\,.}
\end{equation}

\section{Symmetric supergravity theories}\label{symmetric}
As we mentioned in the introduction the definition of a symmetric
supergravity theory is that the scalars span a symmetric coset space
$G_4/H_4$ and that the coupling to the vectors is such that the
dimensionally reduced theory again has a symmetric scalar manifold
$G_3/H_3$ (after dualisation of the vectors). Let us briefly recall
some basic facts about the geometry of symmetric spaces.

Consider a general group $G$ and its associated Lie algebra
$\mathbb{G}$. The symmetric space property is defined through an
involution $\theta$ that respects the Lie bracket. This induces a
natural grading of Lie algebra elements of $\mathbb{G}$ into
``even'' and ``odd''
\begin{equation}
\text{even}:\, \theta (T)= + T\,,\qquad\qquad \text{odd}:\,
\theta(T)=-T\,.
\end{equation}
Because $\theta$ respects the Lie bracket one can show that the even
generators form a subalgebra $\mathbb{H}$. The real group associated
to the Lie algebra $\mathbb{H}$ is accordingly denoted $H$. The
vector subspace of odd generators is denoted $\mathbb{K}$, such that
we can write
\begin{equation}
\mathbb{G}=\mathbb{H} \oplus \mathbb{K}\,.
\end{equation}
This is called the Cartan decomposition of the algebra and $\theta$
is accordingly called the Cartan involution. The properties of the
involution imply, besides $\mathbb{H}$ being a subalgebra, that
\begin{align}
[\mathbb{K},\mathbb{K}]\subset\mathbb{H}\,,\qquad
[\mathbb{K},\mathbb{H}]\subset\mathbb{K}\,.
\end{align}

By letting $\theta$ act through the exponent we can naturally define
the Cartan involution $\theta$ on the level of the group $G$. This
also allows us to define the coset space $G/H$ as spanned by the
elements $x(g)$, with $g\in G$,
\begin{equation}\label{definitionofx}
x(g)=g\theta(g)^{-1}
\end{equation}
such that $\theta(x)=x^{-1}$. For some coordinate system $y$ on
$G/H$ we can parametrize the coset elements as elements of $G$, in
some representation, $L(y)\in G$, where we assume that the isometry
group $G$ acts from the right: $L\rightarrow gL$ and the local
isotropy group from the left $L\rightarrow L h$  \footnote{There is
some difference with the notation we used in previous papers. Often
one uses the symbol $\mathbb{L}$ for the coset representative and
the symbol $L$ for the Lax operator. In this paper we choose $L$ for
the coset representative and $\mathcal{V}$ for the Lax operator.}.
This way the combination (\ref{definitionofx})
\begin{equation}
\mathcal{M}=L\theta (L^{-1})
\end{equation}
is a proper coset element that is invariant under $H$ and transforms
under $G$ as
\begin{equation}
\mathcal{M}\rightarrow g \mathcal{M} \theta(g^{-1})\,.
\end{equation}
We name $\mathcal{M}$ the ``symmetric'' coset matrix and $L$ a coset
representative.

There exists a natural choice of coordinates on symmetric spaces
called horospherical coordinates. In the supergravity literature
this is known under the name ``Borel gauge''. It is a popular
coordinate frame for many reasons, one of them being that these
coordinates are in a simple 1-1 correspondence with the
10-dimensional supergravity degrees of freedom in a canonical basis
\cite{Andrianopoli:1996bq}. This is called a gauge because a
coordinate frame can be seen as a choice for the local isotropy
``compensator'' that keeps the coset representative in a fixed basis
$L=\exp(\phi^i T_i)$. The Borel gauge is then the gauge in which the
$T_i$ are the elements of the Borel subalgebra, which for the
maximally non-compact real form of a simple complex algebra is the
algebra formed by the positive step operators $E_{\alpha}$ and the
Cartan generators $H_i$:
\begin{equation}
L=\exp(\sum_{\alpha}\chi^{\alpha}E_{\alpha})\exp(\tfrac{1}{2}\sum_a\Phi^a
H_a)\,.
\end{equation}
In this equation, we have denoted the positive roots by $\alpha$ and
the simple roots by $a$. The simple roots $a$ have been used to pick
a basis of the Cartan subalgebra, given by the generators $H_a = 2 a
\cdot H/a^2$.
The scalars $\Phi^a$ are referred to as ``dilatons''\footnote{This differs slightly from the supergravity literature where the dilatons are conjugate to the $H_i$ that are mutually othogonal $Tr[H_iH_j]\sim\delta_{ij}$. Therefore the dilatons in our context are non-orthogonal linear combinations of the dilatons in the usual supergravity context.}.  and the
$\chi^{\alpha}$ as ``axions''. In the Riemannian case, where $H$ is
the maximal compact subgroup, the consistency of this gauge is a
consequence of the Iwasawa decomposition, which implies that the
horospherical coordinate system covers the whole manifold. In the
non-Riemannian case, where $H$ is non-compact, the Iwasawa
decomposition doesn't hold everywhere, but it is still a good local
description. Especially in the case of black holes the patch covered
by the horospherical coordinates is the whole physical patch (see
\cite{Chemissany:2009hq, Chemissany:2010zp} for a discussion on
this).

The metric on the symmetric space is induced from the
Cartan--Killing metric of the Lie algebra $\mathbb{G}$ restricted to
the coset part $\mathbb{K}$. This defines the metric on the tangent
space at the origin and using the isometries this defines the metric
everywhere. This proceeds as follows. The left-invariant one-form
$L^{-1}\d L$ is $\mathbb{G}$-valued and can therefore be split
according to the Cartan decomposition
\begin{equation}
L^{-1}\d L= \mathcal{V} + \mathcal{W} \,,\qquad\text{where}\qquad
\mathcal{V}\in\mathbb{K}\,,\qquad \mathcal{W}\in \mathbb{H}\,.
\end{equation}
In particular this means that
\begin{equation}
\mathcal{V}=\tfrac{1}{2}\Bigl(L^{-1}\d L -\theta(L^{-1}\d
L)\Bigr)\,,\qquad \mathcal{W}=\tfrac{1}{2}\Bigl(L^{-1}\d L
+\theta(L^{-1}\d L)\Bigr)\,.
\end{equation}
The metric on $G/H$ is then defined as
\begin{equation}
g_{ij}(\phi)\,\d\phi^i\otimes\d\phi^j=\alpha\text{Tr}(\mathcal{V}
\otimes \mathcal{V})\,,
\end{equation}
where $\alpha$ is a non-zero positive constant that can be chosen at
will and determines the length-scale, or curvature, of the space.
For a given supergravity theory, however, this number cannot be
chosen at will and is fixed by supersymmetry.

The simplest way of understanding the geodesic equations of motion
\begin{equation}\label{geodesic}
\ddot{\phi}^i +\Gamma^i_{jk}\dot{\phi}^j\dot{\phi}^k=0\,,
\end{equation}
is obtained by first rewriting the coset metric in terms of the
matrix one-form $\mathcal{M}^{-1}\d\mathcal{M}$
\begin{equation}\label{Mone-form}
\mathcal{M}^{-1}\d\mathcal{M}=2\theta(L)\mathcal{V}\theta(L^{-1})\,.
\end{equation}
We then find
\begin{equation}
\text{Tr}(\mathcal{V}\otimes\mathcal{V})=-\tfrac{1}{4}\text{Tr}(\d
\mathcal{M}\otimes \d\mathcal{M}^{-1})\,.
\end{equation}
The variation of the latter form of the action gives the following
equations of motion
\begin{equation}
\ddot{\mathcal{M}}-\dot{\mathcal{M}}\mathcal{M}\dot{\mathcal{M}}=0\qquad
\Leftrightarrow \qquad
\boxed{\frac{\d}{\d\tau}\bigl(\mathcal{M}^{-1}\dot{\mathcal{M}}\bigr)=0\,.}
\end{equation}
This equation can be solved trivially
\begin{equation}
\mathcal{M}(\tau)=\mathcal{M}(0)\e^{Q\tau}\,,
\end{equation}
where the matrix $Q$ and $\mathcal{M}(0)$ are constrained by the
condition
\begin{equation}
\theta (\mathcal{M}(\tau))=\mathcal{M}^{-1}(\tau)\,,
\end{equation}
which gives the right number of integration constants (namely 2$n$
for an $n$-dimensional symmetric space).

As explained in \cite{Chemissany:2010zp} one can, in principal,
extract explicit expressions for the various coordinates
$\phi^i(\tau)$ from $\mathcal{M}$ but this is a laborious and
unpractical task. Furthermore we would like to have some insight on
the integrability of the geodesic equations on the level of the
coordinates (scalar fields). For this purpose one introduces the Lax
pair form of the equations of motion. This form is obtained by
differentiation of equation (\ref{Mone-form}), which after some
algebra gives
\begin{equation}
\frac{\d}{\d\tau}[\mathcal{M}^{-1}\dot{\mathcal{M}}]=2\theta(L)\,\bigl(\dot{V}+[\mathcal{W},
\mathcal{V}]\bigr)\,\theta(L^{-1})\,.
\end{equation}
Hence we find the Lax pair form of the geodesic equations of motion
\begin{equation}\label{laxequation}
\boxed{\dot{\mathcal{V}}=[\mathcal{V},\mathcal{W}]\,.}
\end{equation}

From equation (\ref{Mone-form}) we find that the constant matrix $Q$
is related to the Lax matrix $\mathcal{V}$ as follows
\begin{equation}
Q=2\theta(L)\mathcal{V}\theta(L^{-1})\,.
\end{equation}

\section{Integrability}\label{integrability}

Recently the Liouville integrability of the geodesic equations of
motion (\ref{geodesic}) was proven \cite{Chemissany:2010zp}, based
on earlier work by Fr\'e and Sorin \cite{Fre:2009et}. Liouville
integrability is the statement that for a Hamiltonian system of $n$
degrees of freedom we can find $n$ constants of motion $f_i$ that
mutually Poisson commute
\begin{equation}
\{f_i, f_j\}=0\,,
\end{equation}
where the Poisson bracket and the generalised momenta are defined in
the usual way
\begin{equation}\label{pb}
\{f, g\}=\sum_i \frac{\partial f}{\partial \phi^i}\frac{\partial
g}{\partial p_i} - \frac{\partial g}{\partial \phi^i}\frac{\partial
f}{\partial p_i}\,,\qquad
p_i=\frac{\partial\mathcal{L}}{\partial\dot{\phi}^i}\,.
\end{equation}
For the case at hand the generalised momenta are simply
$p_i=g_{ij}(\phi)\dot{\phi}^j$.

Liouville integrability is equivalent to the fact that the
Hamilton-Jacobi problem has a complete solution. This means that we
can find a globally defined, and generically multi-valued,
generating function $W(\phi_i, f_i)$ that describes the following
canonical transformation
\begin{equation}
(\phi^i, p_i)\quad \overset{W}{\longrightarrow} \quad (\phi^i,
f_i)\,,
\end{equation}
such that the new generalised momenta are the conserved quantities.
It then follows from the theory of generating functions that we have
the following \emph{first-order} equations
\begin{equation}
\dot{\phi}^i=g^{ij}\partial_i W\,,
\end{equation}
where $W$ can effectively be seen as a function of the $\phi^i$
since the $f$ are constants. It was observed and emphasized in
\cite{Andrianopoli:2009je} that this is what underlies the
first-order, or fake supergravity, formalism for black holes (see
\cite{Skenderis:2006rr} for earlier remarks for the case of domain
walls and cosmologies).

Consider a vanishing  NUT charge such that the ${\bf Z}$ become
cyclic, which implies that their generalised momenta, denoted ${\bf
P}$, are constants of motion. Then the function $W(\phi^i, f_i)$ can
be written as  \cite{Andrianopoli:2009je}
\begin{equation}\label{WW4P}
W(\phi^i)= W_4(\phi^r, U) + {\bf Z}\cdot {\bf P}\,.
\end{equation}
Where the function $W_4(\phi^r, U)$ does not depend on the ${\bf Z}$
and  describes the generating function for the system with black
hole effective potential. Hence the cyclic property of the ${\bf Z}$
coordinates guarantees that the integrability of the geodesic motion
projects to the integrability of the black hole effective potential
system (\ref{S_b}). As we explain below, projected integral motions
are generically \emph{not} integrable and it is a non-trivial
result when they are.

The integrability of (\ref{S_b}) can also be understood from the
point of view of Poisson commuting constants of motion. For that we
use some properties of the constants of motion for the geodesics
that were constructed in  \cite{Fre:2009et,Chemissany:2010zp}. For
our purposes it suffices to mention that the constants are divided
into three classes, which we denote as
\begin{itemize}
\item $\mathcal{H}_{m}(\mathcal{V})$: the polynomial Lax constants of motion.
\item $\mathcal{H}_r(\mathcal{V})$: the rational Lax constants of motion.
\item $\mathcal{H}_A(Q)$: the  $Q$-constants of motion\,.
\end{itemize}
The constants $\mathcal{H}_m(\mathcal{V}),
\mathcal{H}_r(\mathcal{V})$ are functions of the components of
the Lax operator \emph{along the Borel algebra}. The constants
$\mathcal{H}_A(Q)$ are obtained by replacing the Lax
components along the Borel algebra by the components of $Q$ along
the Borel algebra for those $\mathcal{H}(\mathcal{V})$ that are not Casimirs \cite{Chemissany:2010zp}.
As the names suggest the polynomial constants are polynomial
functions of $\mathcal{V}$. The polynomial Lax constants are
combinations of the $\text{Tr}(\mathcal{V}^N)$ functions, which were
known to Poisson commute \cite{Olive:1983mw}. One of those, namely
$\text{Tr}\mathcal{V}^2$ corresponds to the Hamiltonian.

A crucial property of the constants of motion is that \emph{the
number of Q-constants $\mathcal{H}_A(Q)$ is equal to the number of
$\bold{Z}-$axions that are integrated out}. In fact, when the Taub-NUT charge
vanishes one can replace the $Q$-constants with the electric and
magnetic charges ${\bf Q}$ since they are the same in number and
also mutually commute amongst themselves and with the constants
built from the Lax operator. It is now  easy to show that the
remaining constants $\mathcal{H}(\mathcal{V})$ are commuting
constants of motion for the smaller system. To prove this it is
sufficient to observe that the Lax operator $\mathcal{V}$ only
contains the axions related to the vectors in a total derivative
i.e. as $\dot{\bf{Z}}$. Hence,
\begin{equation}
\mathcal{H}(\mathcal{V})=\mathcal{H}(\mathcal{V})[U,\phi^r; p_U, p_r,{\bf P}]\,.
\end{equation}
This can be seen as a function of the phase space variables of
the truncated system $(U,\phi^r; p_U, p_r)$ since the ${\bf P}$ are
constants. Now it is straightforward to check that the Poisson
bracket relations on the larger phase space imply that the Poisson
bracket on the smaller phase space also vanish.

\section{Toda molecule equations}\label{todamolecules}

The procedure of integrating out the ${\bf Z}$-axions (and the $a$
axion) generalizes the connection between geodesics on symmetric
spaces and Toda molecules \cite{Olshanetsky:1981dk,
Ferreira:1984bi}. The connection between Toda molecules and
geodesics is established by integrating out specific directions in
order to obtain a non-geodesic motion on the truncated space. Let us
here apply this specifically to the case of coset spaces relevant to
describing black holes in $D=4$ and stick to the case of maximally
split cosets. These results also extend to the case of non-maximally
split cosets and sigma models not specific to $D=4$ black holes.

Consider the sigma model in $D=4$, denoted $G_4/H_4$, where $H_4$ is
the maximal compact subgroup of $G_4$. Dimensional reduction then
gives a coset $G_3/H_3$. $\mathbb{G}_3$ contains the two disjoint
subalgebras, $A_1$ and $\mathbb{G}_4$. The $A_1$ subalgebra is built
from the extra Cartan generator $H_{\beta_{0}}$ that is generated from
the reduction from four to three dimensions and $\beta_{0}$ denotes the
corresponding simple root. The adjoint of $\mathbb{G}_3$ decomposes
as follows under $A_1\times \mathbb{G}_4$
\begin{equation}
\mathbb{G}_3\rightarrow (1, \mathbb{G}_4)\oplus (3, 1)\oplus (2, \bf
R)\,,
\end{equation}
where $\bold{R}$ is some subspace whose representation we do not specify
any further.

We can divide the positive roots $\alpha$ in three classes, depending on
their behavior under $A_1$, which can be read off from the weight
the generator has w.r.t. $H_{\beta_{0}}$
\begin{enumerate}
\item $E_{\beta_{0}}$ has weight two under $H_{\beta_{0}}$.
\item the positive roots $E_{\gamma}$, that constitute the representation ${\bf
R}$, have weight one under $H_{\beta_{0}}$. \item the positive roots in
$\mathbb{G}_4$,  denoted $E_{\beta}$, are orthogonal to $\beta_{0}$
(and thus have no weight under $H_{\beta_{0}}$).
\end{enumerate}
Because of the grading structure, one can show via the Jacobi
identity, that the positive roots $\gamma$ that build the
representation ${\bf R}$ obey
\begin{equation}\label{Heisenberg}
[E_{\gamma_1}, E_{\gamma_2}]=\mathbb{C}_{\gamma_1 \gamma_2}
E_{\beta_{0}}\,.
\end{equation}
In the context of dimensional reductions to three dimensions one
should think of the roots and the corresponding axions $\chi$ as
follows. The $\chi^{\beta_{0}}$ corresponds to the Hodge-dual of the KK
vector that arises in going from four to three dimensions. The
$\chi^{\gamma}$ are the axions ${\bf Z}$ that come from the vector
potentials in three dimensions and the $\chi^{\beta}$ are the
axionic scalars already present in four dimensions. Their grading
structure can be read off from the power of $\e^{-2U}$ that is in
front of their kinetic terms (\ref{S_a}), since the scalar $U$ is
the dilaton that multiplies $H_{\beta_{0}}$ in the coset representative.

This grading structure allows us to define the
integrations/truncations we are looking for: 1) The truncation of
$\chi^{\beta_{0}}$, 2) The truncation of $\chi^{\beta_{0}},\chi^{\gamma}$
and 3) The truncation of all axions $\chi^{\alpha}.$

The direction related to the root $\beta_{0}$ can be integrated out, as
it appears shift symmetric. This means that
\begin{equation}
\e^{-4U}(\dot{a} + {\bf Z}^T \mathbb{C}\dot{{\bf Z}})=n\,,
\end{equation}
where $n$ is a constant proportional to the Taub-NUT charge. After
this integration the effective action is given by
\begin{equation} \label{TNintegrated}
S= \int d\tau\left( \dot{U}^{2} +
\tfrac{1}{2}G_{rs}\dot{\phi}^{r}\dot{\phi}^{s} +
\tfrac{1}{2}e^{2U}\bold{Q}^{T}\mathcal{M}\bold{Q} +
\tfrac{1}{4}e^{4U}n^{2}\right),
\end{equation}
where, in this case,  the physical electric and magnetic charges take the form
\begin{equation}\label{emcharges}
\bold{Q}=\mathbbm{C}\bold{P}-\frac{1}{2}\bold{Z}n\,.
\end{equation}
where the momentum $\bold{P}$ contains $\dot{\bold{Z}}$ and is
defined as the momentum conjugate to $\bold{Z}$.  This effective action can be seen as an effective action of the variables $U, \phi^r, \bold{Z}$ or of the smaller set of variables where the $\bold{Z}$ are integrated out because the $\bold{Q}$  in (\ref{TNintegrated})  are constants of motion. In case we include the variables $\bold{Z},$ integrability is again inherited from the geodesic as follows. The Taub-Nut charge $n$ coincides with one of the Hamiltonians of the geodesic motion \cite{Chemissany:2010zp}, and furthermore the remaining Hamiltonians do not depend on $a$  and only depend on $\dot{a}$  via the TN charge $n$. Therefore the remaining Hamiltonians are the required Hamiltonians necessary for Liouville integrability. We expect the integrability of the geodesic to also project down to the integrability of this effective action regarded as a function of $U$ and $\phi^r$ only. However, this has not yet been shown because some of the Hamiltonians depend explicitly on $\bold{Z}$. This is currently under investigation.

From the effective action (\ref{TNintegrated}) we notice that only
when $n=0$ the ${\bf Z}$ axions appear shift symmetric. This is
because the Heisenberg algebra (\ref{Heisenberg}) becomes Abelian.
The shift symmetry implies that we can integrate out the
$\chi^{\gamma}$ for any value of their velocities. For the black
hole case this leads to the action with the black hole effective
potential (\ref{S_b}). As we explained in the previous section the
Liouville integrability of the geodesic motion carries over to this
truncated system.

Finally, one can integrate out all the axions $\chi^{\alpha}$ but this is not
possible for generic velocities. It is explained in
\cite{Ferreira:1984bi} when this is possible and we recall this in
what follows since this is how the link with conventional Toda
molecules appears. We also immediately extend the analysis by
allowing an isotropy group $H$ which is not necessarily compact, as
for the case at hand. We consider a Cartan involution that is
defined as follows
\begin{equation}
\theta (H_i)=-H_i\,,\qquad \theta
(E_{\alpha})=-\epsilon_{\alpha}E_{-\alpha}\,,
\end{equation}
where $\epsilon_{\alpha}$ is a specific sign. For the case of
dimensional reductions from four to three we have the conventional
choice, $\epsilon_{\alpha}=+1$, when we reduce over a spacelike
direction. This leads to a Riemannian coset for which $H$ is the
maximal compact subgroup of $G$. For dimensional reductions over
time we have $\epsilon_{\alpha}=(-1)^{\alpha\cdot\beta}$, such that
for the roots $\gamma$ we have the unconventional sign. We will
however keep the sign $\epsilon_\alpha$ arbitrary, since the
following argument works for more general Cartan involutions.

For the purpose of integrating out we have to consider the
(pull-back of the) one-forms $L^{-1}\d L$ and
$\mathcal{M}^{-1}\d\mathcal{M}$, built from the coset representative
in the Borel gauge. We find \footnote{One should use the Hadamard
lemma : $\e^T S \e^{-T}=\e^{adj T} S$.}
\begin{equation}\label{LdL}\Omega=
L^{-1}\dot{L} = \tfrac{1}{2}\sum_a \dot{\Phi}^a H_a + \sum_{\alpha}
V_{\alpha}\e^{-\tfrac{1}{2} \sum_a K_{\alpha a}\Phi^a}E_{\alpha}\,,
\end{equation}
where $V_{\alpha}$ are functions of the $\chi$ and $\dot{\chi}$
defined via
\begin{equation}
\sum_{\alpha}
V_{\alpha}E_{\alpha}=\e^{-\sum_{\alpha}\chi^{\alpha}E_{\alpha}}\,\frac{\d}{\d\tau}\,
\e^{\sum_{\alpha}\chi^{\alpha}E_{\alpha}} \,,
\end{equation}
and $K$ is defined as
\begin{equation}
K_{\alpha a}=2\frac{\alpha\cdot a}{ a^2}\,.
\end{equation}
The expression for $\mathcal{M}^{-1}\dot{\mathcal{M}}$ is more
involved and we write it as
\begin{equation}\label{M-1dotM}
\mathcal{M}^{-1}\dot{\mathcal{M}}=\e^{-\sum_{\alpha}\epsilon_{\alpha}\chi^{\alpha}E_{-\alpha}}\Bigl(
\sum_{\alpha} V_{\alpha}\e^{- \sum_a K_{\alpha a}\Phi^a}E_{\alpha} +\sum_a
\dot{\Phi}^a H_a +\sum_{\alpha}T_{\alpha}E_{-\alpha}
\Bigr)\e^{\sum_{\alpha}\epsilon_{\alpha}\chi^{\alpha}E_{-\alpha}}\,,
\end{equation}
where the $T_{\alpha}$ are functions of the axions and their
derivatives, defined as follows
\begin{equation}
\sum_{\alpha}T_{\alpha}E_{-\alpha}=\frac{\d}{\d\tau}
\bigl(-\e^{-\sum_{\alpha}\epsilon_{\alpha}\chi^{\alpha}E_{-\alpha}}\bigr)\,
\e^{\sum_{\alpha}\epsilon_{\alpha}\chi^{\alpha}E_{-\alpha}}\,.
\end{equation}

If we use the standard Cartan-Weyl commutation relations it is easy
to see that the coefficient in front of the highest positive root
$E_{\alpha}$-term in (\ref{M-1dotM}) is simply
$V_{\alpha}\e^{-\sum_aK_{\alpha a}\Phi^a}$. Since
$\mathcal{M}^{-1}\dot{\mathcal{M}}$ is constant, so is that term.
If we therefore take it to vanish initially, it vanishes at all
times. Clearly we can subsequently repeat this for the new highest
root and so on such that we end up with the simple roots only. Then
one can similarly infer that the coefficients in front of the $E_a$
are simply $V_{a}\e^{-\sum_bK_{a b}\Phi^b}$. Hence they are constant
and we denote these constants by $C_a$. This can be seen as a way to
integrate out the dependence on the axions and their derivatives
since the only place where they appeared, after all the truncations
made sofar, is in $V_a$. But we can write $V_a$ in terms of the
dilatons
\begin{equation}
V_a=C_{a}\e^{\sum_bK_{a b}\Phi^b}\,.
\end{equation}
If we plug this in the definition of $\mathcal{V}$ and
$\mathcal{W}$, we find (after the truncations we made)
\begin{align}
\mathcal{V}&=\tfrac{1}{2}\sum_a\dot{\Phi}^a H_a
+\tfrac{1}{2}\sum_{a} C_a \e^{\tfrac{1}{2}\sum_bK_{ab}\Phi^b}(E_a +
\epsilon_a E_{-a})\,,\\
\mathcal{W}&= \tfrac{1}{2}\sum_{a} C_a
\e^{\tfrac{1}{2}\sum_bK_{ab}\Phi^b}(E_a - \epsilon_a E_{-a})\,.
\end{align}
The Lax equation (\ref{laxequation}) then leads to
\begin{equation}\label{dtodaeq}
\boxed{\ddot{\phi^a}=-\epsilon_a C_a^2
\e^{\tfrac{1}{2}\sum_b{K_{ab}\Phi^b}}\,.}
\end{equation}
Either by choosing initial conditions or by shifting the $\Phi^a$ we
can take $C_a^2=1$ and this then defines the Toda molecule
equations, when $\epsilon_a=1$, i.e., when the isotropy group is
compact. When not all the $\epsilon_a$ are equal to $1$ we find a
generalization\footnote{A similar generalisation appears in
\cite{Feher:1995gh}, where also the relation with geodesic curves
was used.} of the Toda molecule equations which still inherits the
integrability of the underlying geodesic motion, as can easily be
shown using the same arguments of \cite{Ferreira:1984bi,
Olive:1983mw}. These arguments are based on the existence of an
operator $\mathbbm{P}$ that obeys a set of relations, called the
fundamental Poisson relations. Essentially, the operator
$\mathbbm{P}$ allows one to rewrite the Poisson bracket in terms of
a Lie algebra commutator. The existence of the operator
$\mathbbm{P}$ makes it manifest that the quantities $\textrm{Tr}
\mathcal{V}^{N}$ are in involution
\begin{equation}\{\textrm{Tr} \mathcal{V}^{N},\textrm{Tr} \mathcal{V}^{M}\}=0.
\end{equation}
As the number of independent quantities $\textrm{Tr}
\mathcal{V}^{N}$ is given by the rank of the Lie algebra, this
establishes Liouville integrability of the Toda molecule equations.
The operator $\mathbbm{P}$ constructed in
\cite{Ferreira:1984bi,Olive:1983mw} has a uniform structure, that
depends on the root system of the underlying algebra
$\mathbbm{G}_{3}$. We have checked explicitly that the same operator
$\mathbbm{P}$ also establishes the Liouville integrability of the
$\epsilon$-generalised Toda molecule equations, as suggested in
\cite{Ferreira:1984bi}.


\subsection*{An example}

Consider the Einstein--Maxwell--dilaton action
\begin{equation}\label{EMD}
S=\int\sqrt{|g|}\Bigl(\tfrac{1}{2}\mathcal{R}-\tfrac{1}{2}(\partial\phi)^2
-\tfrac{1}{4}\rme^{\sqrt{6}\phi}F^2 \Bigl)\,.
\end{equation}
The action (\ref{EMD}) is just the circle reduction of pure gravity
in $d=5$.  The reduction over time leads to the
$\SL(3,\Real)/\SO(2,1)$-coset. The black hole solutions of this
theory have first been considered in \cite{Dobiasch:1981vh}. After
truncation of the Taub-NUT direction the $\SL(3)$-sigma model becomes
\begin{equation}
S_a=\int \d\tau \,\, \left\{\dot{U}^2 + \tfrac{1}{2}\dot{\phi}^2
-\tfrac{1}{2}\e^{-2U+\sqrt{6}\phi}(\dot{Z}^1)^2 -
\tfrac{1}{2}\e^{-2U-\sqrt{6}\phi}(\dot{Z}^2)^2\right\} \,,
\end{equation}
If we subsequently integrate out the ${\bf Z}$ axions we find the
action
\begin{equation}\label{action}
S_a=\int \d\tau \,\,\left\{\dot{U}^2 + \tfrac{1}{2}\dot{\phi}^2 +
e^{2U}V(\bold{Q},\phi)\right\},
\end{equation}
where the positive definite potential $V(\bold{Q},\phi)$  has the following form
\begin{equation}
V=\frac{1}{2}\bold{Q}^{T}\mathcal{M}_{4}\bold{Q}=\frac{1}{2}\bold{Q}^{T}
\begin{pmatrix} e^{\sqrt{6}\phi}&0 \\ 0&e^{-\sqrt{6}\phi} \end{pmatrix} \bold{Q},\qquad
\bold{Q}=(m,e),
\end{equation}
with
\begin{equation}
\bold{P} =
\Big(-e^{-2U+\sqrt{6}\phi}(\dot{Z}^{1}),-e^{-2U-\sqrt{6}\phi}(\dot{
Z}^{2})\Big), \qquad \bold{Q}=\mathbbm{C}\bold{P}\,.
\end{equation}

Let us now show how the generalized Toda molecule equation
(\ref{dtodaeq}) reproduces the equations of motion following from
(\ref{action}), namely,
\begin{equation}\label{Uphi}\ddot{U}=e^{2U}V,\qquad \ddot{\phi}=e^{2U} \frac{\partial V}{\partial \phi}.\end{equation}
The only thing we need is the Cartan matrix for $\SL(3)$
\begin{equation}
K=\begin{pmatrix}2&-1\\-1&2\end{pmatrix}.
\end{equation}
Thus, the Toda molecule equation  (\ref{dtodaeq}) can be written out
to yield
\begin{eqnarray}\label{sl3toda1}
\ddot{\Phi}^{1}&=&C_{1}^{2}e^{\frac{1}{2}(2\Phi^{1}-\Phi^{2})},\\
\ddot{\Phi}^{2}&=&C_{2}^{2}e^{\frac{1}{2}(-\Phi^{1}+2\Phi^{2})}\,.\label{sl3toda2}\,
\end{eqnarray}
One can go from equations (\ref{sl3toda1})-(\ref{sl3toda2}) to
equations (\ref{Uphi}) by simply performing the following field
redefinitions
\begin{eqnarray}\Phi_{1}&=&4(U+\frac{1}{\sqrt{6}}\phi)\,,\\
\Phi_{2}&=&4(U-\frac{1}{\sqrt{6}}\phi)\,,
\end{eqnarray}
together with the identification $C_{1}=2 m,$  $C_{2}=2 e$.

Few other examples exist for which it was known that the black hole
equations of motion reduce to a Toda system, see
e.g.~\cite{Monni:1995vu, Astefanesei:2006sy}.

\section{Discussion}\label{discussion}

The fact that solutions to the system
\begin{equation}\label{potential}
\mathcal{L}=\tfrac{1}{2}g_{rs}\dot{\phi}^r\dot{\phi}^s - V(\phi^r)
\end{equation}
can sometimes be seen as a subset of solutions to a free system of
    \emph{more} degrees of freedom\footnote{Since the configuration space of the geodesic has more dimensions this is different from the Maupertuis--Jacobi principle. }
\begin{equation}\label{free}
\mathcal{L}=\tfrac{1}{2}g_{ij}\dot{\phi}^i\dot{\phi}^i
\end{equation}
finds a concrete realisation for black hole equations of motion in supergravity.  Note that (\ref{free}) is not unique since there can exist multiple extensions of the configuration space such that a truncation of geodesics on the larger space coincide with the equations of motion coming from (\ref{potential}). The
reformulation of (\ref{potential}) as (\ref{free}) is useful when
(\ref{free}) exhibits more symmetries. This is the case for black
hole solutions in symmetric supergravity theories and the
reformulation (\ref{free}) establishes the full integrability of
(\ref{free}) and (\ref{potential}).

The equivalence between (\ref{free}) and (\ref{potential}) is more general than black holes in supergravity, see for instance  \cite{Townsend:2004zp} in the context of cosmological solutions. In fact it applies to general ``brane'' solutions\footnote{In this language we regard a FRW cosmology as a S-brane solution with one transversal direction being time. Just like a stationary domain wall where the transversal direction is spacelike. }. This can easily be understood due to the fact that there are two
ways to describe the effective action for brane solutions
\cite{Bergshoeff:2008zza}. Either one reduces the brane over its
flat worldvolume and then the effective action is a geodesic motion
\cite{Bergshoeff:2008be} or one reduces the brane over that part of
the transversal space that does not include the ``radial
direction'', such as a sphere for spherical brane solutions. This
maps the brane solution to a domain wall solution of a gauged
supergravity in a lower dimension. The latter effective action is of
the kind (\ref{potential}). As explained in
\cite{Bergshoeff:2008zza} this is equally valid for time-dependent
brane solutions, such as $S$-branes, which get mapped to cosmological
solutions in the lower-dimensional theory, which are again described
by an action of the form (\ref{potential}). This gives a higher-dimensional interpretation to the observations made in
\cite{Townsend:2004zp}.

The above considerations lead us to conjecture that domain wall (and
cosmology) effective actions in gauged maximal supergravity might be
integrable because of a hidden geodesic motion on a larger target
space, maybe in some Kac-Moody sigma model related to the
$E_{10}$  or $E_{11}$ algebra. This picture comes about when one assumes that all
domain walls (and cosmologies) in maximal supergravity have some kind of brane origin
in 10 or 11 dimensions. Given the fact that the standard branes (and
$S$-branes) in 10 and 11 dimensions can be found as solutions to the
$E_{10}$ or $E_{11}$-sigma model
\cite{Englert:2004it,Kleinschmidt:2005gz} it makes perfect sense to
conjecture the hidden geodesic origin of all domain walls (or
cosmologies) of maximal gauged supergravity.

\section*{Acknowledgements}
We like to thank Mario Trigiante for various valuable discussions on this topic.
W.C. is supported in part by the Natural Sciences and Engineering
Research Council (NSERC) of Canada. T.V.R. is supported by the
G\"oran Gustafsson Foundation. W.C. also likes to thanks the university of Granada for its hospitality during the last stages of this work.

\bibliography{toda}
\bibliographystyle{utphysmodb}

\end{document}